# Angular Clustering of Millimeter-Wave Propagation Channels with Watershed Transformation

Pengfei Lyu, Aziz Benlarbi-Delaï, Zhuoxiang Ren, *senior member, IEEE,*
and Julien Sarrazin, *senior member, IEEE*

*Abstract*—An angular clustering method based on image processing is proposed in this paper. It is used to identify clusters in 2D representations of propagation channels. The approach uses operations such as watershed segmentation and is particularly well suited for clustering directional channels obtained by beam-steering at millimeter-wave. This situation occurs for instance with electronic beam-steering using analog antenna arrays during beam training process or during channel modeling measurements using either electronic or mechanical beam-steering. In particular, the proposed technique is used here to cluster two-dimensional power angular spectrum maps. The proposed clustering is unsupervised and is well suited to preserve the shape of clusters by considering the angular connection between neighbor samples, which is useful to obtain more accurate descriptions of channel angular properties. The approach is found to outperform approaches based on K-Power-Means in terms of accuracy as well as computational resources. The technique is assessed in simulation using IEEE 802.11ad channel model and in measurement using experiments conducted at 60 GHz in an indoor environment.

*Index Terms*—propagation channel angular clustering, millimeter wave, watershed transformation, 2D angular measurements at 60 GHz

## I. INTRODUCTION

RECENT years have witnessed a growing interest in millimeter waves communications [1]. The spectrum congestion in the lower part of the spectrum and the ever-growing need of higher data rates incited the telecommunication actors to assess the suitability of mm-wave frequency bands to support Gbps wireless communications. During the last decade, several standards have been proposed to operate high data rate communications between devices for Wireless Personal Area Network (WPAN), e.g. IEEE 802.15.3, or Wireless Local Area Network (WLAN), e.g., IEEE 802.11ad and more recently IEEE 802.11ay. This trend has been even more emphasized recently with the advent of 5G. The 3GPP release 15 has defined the use of bandwidth in the 24.25-40 GHz range [2] and the release 17 is considering frequencies in the 52.6-71 GHz spectrum, including the 60 GHz license-free band [3]. This profound change in the network infrastructure is however challenging due to the number of specificities that are inherent to operating at such high frequencies. Indeed, mm-wave communications typically use several antennas to achieve array gain in order to mitigate free-space attenuation in budget link [4]. High gain antennas exhibit narrow beams thereby leading to sparse multipath illumination [5].

In scenarios where a base station needs to address mobile users, beamforming precoding or beam steering techniques are required. To assess such communications, one needs directional channel models with realistic distributions of Angle-of-Arrival (AoA) and/or Angle-of-Departure (AoD) [6]. Channel modeling is classically performed by fitting distributions onto features extracted from channel measurements. To obtain AoA/AoD information, full-digital antenna arrays [7] or synthetic array [8], [9] are usually employed and algorithms such as MUSIC [9], [10], SAGE [6], [8], [11], [12], or CLEAN [13], [14] can be used to estimate power, direction of arrival, and Time-of-Arrival (ToA) of multipath components. The channel can be then represented as a discrete data set of features such as power in a 2D plane (azimuth and elevation angles [15] or angle and ToA [8], [11]) or even a 3D plane (both angles and ToA [6]). Each sample in this discrete data set can be modeled by a plane wave. This representation is then clustered and probability density function (PDF) are fitted to describe the behavior of inter- and/or intra-clusters features, whether in angular domain or in time domain [16].

In mm-waves, analog antenna arrays are typically preferred over full-digital architectures or synthetic array to perform outdoor channel measurements in order to be able to benefit from the array gain before the analog-to-digital conversion of the baseband signal. The procedure is then to scan the whole angular range thanks to beam steering (either electronic [17] or mechanical [16]) or beam switching [17], and therefore obtain a quasi-continuous channel representation in the angular/frequency domain. The angular accuracy of this representation depends on the beamwidth of the antenna array

Manuscript received August 13, 2010; revised February 14 and June 14, 2021; accpepted August 19, 2021. This work was supported in part by National Key R&D Program of China under Grant 2017YFB020350, National Science Foundation of China under Grant 61501454. This work was performed within NOVIS60 project supported by CEFPRA (Indo-French Center for the Promotion of Advanced Research).

Pengfei Lyu is with the Institute of Microelectronics, Chinese Academy of Sciences, 100029 Beijing, China; School of Microelectronics, University of Chinese Academy of Sciences, 100049 Beijing, China; CNRS, Laboratoire Génie Electrique et Electronique de Paris, Sorbonne Université, 75005 Paris, France and Université Paris-Saclay, Centrale Supélec, 91192, Gif-sur-Yvette, France (e-mail: lvpengfei@ime.ac.cn; pengfei.lyu@sorbonne-universite.fr).

Julien Sarrazin (e-mail: julien.sarrazin@sorbonne-universite.fr), Aziz Benlarbi-Delaï (e-mail: aziz.benlarbi_delai@sorbonne-universite.fr) and Zhuoxiang Ren (zhuoxiang.ren@sorbonne-universite.fr) are with CNRS, Laboratoire de Génie Electrique et Electronique de Paris, Sorbonne Université, 75005 Paris, France and Université Paris-Saclay, Centrale Supélec, 91192, Gif-sur-Yvette, France.

Color versions of one or more of the figures in this communication are available online at https://doi.org/10.1109/TAP.2021.xxx.

Digital Object Identifier 10.1109/TAP.2021.xxx.

and the angular step size. Based on this representation, regular techniques can still be applied in time domain to obtain the ToA discrete data set while high-resolution algorithms such as MUSIC have been adapted to operate in such a beam space representation [18] to estimate AoA/AoD and thus form the discrete data set [19] onto which classical channel modeling procedures typically used in lower part of spectrum, including clustering and PDF fitting, can be similarly applied [16].

Identifying cluster shapes in time domain is efficiently performed using a priori knowledge, typically an exponential decay with increasing delay [20]. While this assumption is physically quite realistic, doing so to find clusters in the angular domain, i.e., the power angular spectrum (PAS), is not optimal as their shapes heavily depend on the scenario and the environment. For instance, intra cluster angular distributions have been variously modeled in the literature by an exponential decay in [21], a Laplacian distribution in [16], [22], and a Von Mise distribution in [23]. So, there is still a need of unsupervised clustering methods that preserves real cluster shapes to more accurately describe channel features [24]. This is especially important to assess techniques that are sensitive to PAS (see, e.g. [25], [26] for AoA/AoD estimation or [27] for multi-user power allocation in massive MIMO context) or that use PAS as a priori knowledge (see, e.g. [28] for beam training improvement or [29], [30] for AoA/AoD estimation). Furthermore, the clustering method should be fast enough since to obtain statistically meaningful results, a large number of channels is to be analyzed [24].

Most of the current propagation channel model in the literature use K-Power-Means (KPM) algorithm as the clustering method [31]. KPM algorithm is a modified version of the general K-Means [32] clustering method. K-Means aims to minimize the sum of the error between the centroid and the components in all of the clusters, by minimizing the average Euclidean distance between data points within a cluster and the mean of the cluster while KPM minimizes the sum of power-weighted distances of parameter points to the centroid associated with the parameter point [31], [33]. K-Means-based cluster analysis has intrinsic weaknesses. Firstly, the number of clusters has to be assumed before the operation. This implies to fix the number of clusters based on visual inspection [34] or to use some automatic detection process based on a priori knowledge [35]. However, it has been observed in [36] that when different clusters exhibit different statistics, the automatic detection may fail. Another approach is an incremental search for that appropriate number, using convergence threshold such as cluster power with respect to total power [8], [37] or graphical-based metrics such as silhouettes [38] for instance, albeit at the expense of higher computational resources. Secondly, inappropriate initial clusters lead to local minima. To solve the initializing problem, the K-Means++ algorithm [39] was introduced to initialize the centroid of the cluster randomly. Thirdly and importantly for this work, K-means-based methods are among globally clustering methods and therefore treats all features equally, regardless the actual correlation among them. This third problem led to the fact that the channel clusters does not reflect accurately the channel impulse responses (CIR) exponential decrease with time in [20] and the CIR had to be fitted with a priory known exponential function to solve this issue [20]. However, a priory functions destroy the unsupervised nature of K-Means. While approaches have been proposed to weight differently features in K-means, the weighting factors are often found empirically and are scenario- dependent [40].

To address these shortcomings, this paper introduces the use of image signal processing techniques to obtain an efficient and unsupervised approach for clustering channels in angular domain that considers the connection between neighbor samples. Indeed, not taking into account the location correlation may result in faraway samples being grouped into the same angular cluster which jeopardizes preserving its actual shape. The idea is not to work on the extracted features but directly on a quasi-continuous channel representation in 2D, namely, the PAS along elevation and azimuth angles. This allows for the use of a set of morphological operations borrowed from image processing to identify clusters while preserving their angular shape. In particular, a watershed segmentation is performed and the potential of this approach is assessed at 60 GHz with simulations using the IEEE 802.11ad channel model and with measurements in an indoor scenario. The section II describes the considered scenario and the channel representation used in this paper. The proposed clustering algorithm is introduced in section III while the performance is assessed in section IV by comparing the performance with KPM and a modified version of KPM. The clustering method is validated with measurements in section V. Finally, section VI draws conclusions and gives some perspectives.

## II. PROBLEM STATEMENT

To illustrate the clustering technique proposed in this paper as well as to assess its performance, the scenario depicted in Fig.1(a) is considered. An omnidirectional TX antenna and a directional RX antenna with a beamwidth ranging from $5°$ to $29°$ scan the 2D angular space with a step fixed to $1°$.

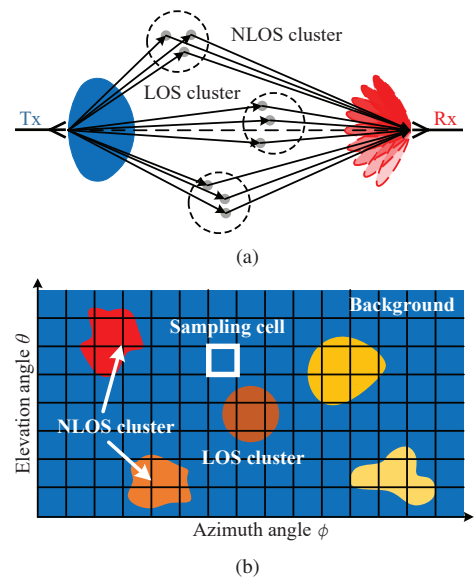

Fig. 1. The generation of power angular spectrum (PAS): (a) angular scanning; (b) resulting 2D PAS grid in azimuth-elevation plane

This situation typically occurs in mm-wave while conducting channel modeling experiments with directional antennas [16], [17] or during beam-training process in mm-wave communications to find the strongest link, i.e., the strongest cluster, between a transmitter and a receiver [41], [42]. A 2D PAS is therefore obtained such as shown in Fig.1(b). This quasi-continuous channel representation forms an image of pixels (i.e., the sampling cell) whose size depends on the angular step size and whose intensity, in grayscale, depends on the channel power in that particular direction. Imaging processing can therefore be applied to such PAS representation to perform efficient clustering. To generate such 2D maps as a data set onto which the clustering is performed, the IEEE 802.11ad directional channel model is used throughout this paper [43], [44] for simulations. The IEEE 802.11ad is a standard for indoor wireless communication in the 60 GHz band. Its channel model is a time and angular clusters-based model. The scenario considered in this paper is the conference room.

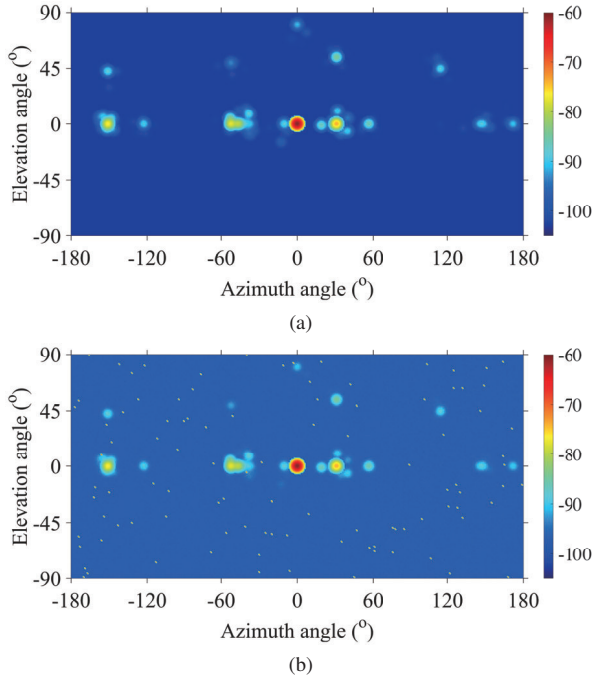

Fig. 2. Angular characteristics generated with IEEE 802.11ad channel model for both azimuth and elevation in dB: (a) original PAS generated by channel model noiseless; (b) PAS with AWGN and random speckles.

The PAS is obtained by the following formula:

$$PAS = \int_0^T |g_t(\theta,\phi) g_r(\theta,\phi) h(t,\theta,\phi) + n(t)|^2 dt \quad (1)$$

where $h(t,\theta,\phi)$ is the CIR. $g_t(\theta,\phi)$ and $g_r(\theta,\phi)$ are the antenna gains of the transmitter and receiver with respectively. $n(t)$ is the thermal noise, Added White Gaussian Noise (AWGN). An example of noise-free PAS generated with a 5° Rx beamwidth is shown in Fig.2(a). The Line-Of-Sight (LOS) component appears at $\theta = \phi = 0°$ while clusters at other angles are due to reflections and diffractions within the environment. In actual measurements, in addition to thermal noise, spatial speckles are also present. They widely appear in images obtained by synthetic aperture radar (SAR) [45], laser [46], and millimeter wave [47]. Speckles occur because of the stochastic coherent combination of a number of independent waves scattered in the environment. To model this effect, 100 speckles uniformly distributed in the angular plane are generated with identical power, equal to the PAS maximum power. This has been found empirically relevant with the experiments conducted and presented in section V. The resulting PAS is shown in Fig.2(b). where an AWGN of SNR = 20 dB is also added. Compared with the original PAS in Fig.2(a), the background power is now higher and exhibits a weak fluctuation. Adding an AWGN in the CIR in (1). results in a spatial noise that follows a biased non-Gaussian distribution leading to a PAS mean SNR of 21.03 dB with a standard deviation 4 dB. Speckles occupy single pixels. Both AWGN and speckles are to be removed to perform accurate clustering and using a simple threshold does not perform generally well. Next section shows how image-processing filtering techniques can remove them efficiently before performing clustering.

## III. CLUSTERING METHOD

### A. Morphological Operations for Watershed

Mathematical morphology (MM) is an imaging processing method to extract information based on set theory and lattice theory. A grayscale image is regarded as a function $f(x)$ that maps a set of 2D coordinate $x$ (pixel position) to a 3D surface extended to the third dimension (pixel value). In the situation in Fig.1(b), the variable $x$ is the discrete angle vector $(\phi,\theta)$, where $\phi$ is the azimuth angle and $\theta$ is the elevation angle. The function $f(x)$ maps the whole angular plane to the received power, $f(x): X^2 \to Y$. $x \in X^2$ and $X^2$ is a 2D coordinate set of the whole angular plane:

$$X^2 = \left\{ x = (\phi,\theta) \,|\, \phi \in [-\pi,\pi], \theta \in \left[-\frac{\pi}{2},\frac{\pi}{2}\right] \right\} \quad (2)$$

The generated 3D space is defined with a set $X^2 \times Y$.

$$X^2 \times Y = \left\{ (\phi,\theta,P) \,|\, \phi \in [-\pi,\pi], \theta \in \left[-\frac{\pi}{2},\frac{\pi}{2}\right], P \in \mathbb{R}_+ \right\} \quad (3)$$

The idea of MM is remodeling the 3D space of an image with local functions, which are called structuring elements. A structuring element is also a mapping to angular plane, $g(x): X_g^2 \to Y$, where $X_g^2 \subseteq X^2$. The reconstruction performed in this paper is achieved with some basic operations that are defined below.

*Operation* 1: **dilation** [48] $f \oplus g : X^2 \oplus X_g^2 \to Y$ is used to extend the local spaces. It extracts the supremum of the sum of $f$ and $g$ at each sliding position of $f$:

$$(f \oplus g)(x) = \sup \{f(x - x') + g(x')\} \quad (4)$$

*Operation* 2: **erosion** [48] $f \ominus g : X^2 \ominus X_g^2 \to Y$ is used to shrink the local image spaces. It extracts the infimum of the difference of f and g at each negatively sliding position of $f$:

$$(f \ominus g)(x) = \inf \{f(x + x') - g(x')\} \quad (5)$$

*Operation* 3: **opening** [48] removes bright peaks that are small in size and break narrow connections between two bright peaks with dilation $\oplus$ and erosion $\ominus$:

$$f \circ g = (f \ominus g) \oplus g \quad (6)$$

*Operation* 4: **closing** [48] preserves small peaks which are brighter than the background and fills the small gaps between bright peaks with dilation $\oplus$ and erosion $\ominus$:

$$f \cdot g = (f \oplus g) \ominus g \tag{7}$$

*Operation* 5: **Euclidean distance transformation** [49] $d(\boldsymbol{x}, \boldsymbol{x}')$ is an operation for a binary image. It assigns the value of each pixel $\boldsymbol{x}$ in a subset $A$ of the whole image with the Euclidean distance between $\boldsymbol{x}$ and the nearest nonzero pixel $\boldsymbol{x}'$ inside a given connected domain $A$:

$$d(\boldsymbol{x}, \boldsymbol{x}') = \inf\left\{\sqrt{\|\boldsymbol{x}_i - \boldsymbol{x}'_j\|^2} \mid A \subset X^2, \boldsymbol{x}, \boldsymbol{x}' \in A, P_A \neq 0\right\} \tag{8}$$

*Operation* 6: **geodesic distance** [50] $d_A(\boldsymbol{x}, \boldsymbol{x}')$ is also on the plane $X^2$. It is the length of the shortest path linked two pixels $\boldsymbol{x}$ and $\boldsymbol{x}'$ in a connected space $A$ constructed by neighbor pixels with identical intensity level.

$$d_A(\boldsymbol{x}, \boldsymbol{x}') = \inf\left\{\sqrt{\|\boldsymbol{x}_i - \boldsymbol{x}'_j\|^2} \mid A \subset X^2, \boldsymbol{x}, \boldsymbol{x}' \in A\right\} \tag{9}$$

*Operation* 7: a **geodesic ball** [51] $\Omega_A(\boldsymbol{x}, \lambda)$ with a center $\boldsymbol{x}$ and radius $\lambda$ is defined as a domain set $\boldsymbol{x}'$ whose geodesic distance $d_A(\boldsymbol{x}, \boldsymbol{x}')$ to x is not larger than $\lambda$:

$$\Omega_A(\boldsymbol{x}, \lambda) = \{\boldsymbol{x}' \in A \mid A \subset X^2, d_A(\boldsymbol{x}, \boldsymbol{x}') \leq \lambda\} \tag{10}$$

*Operation* 8: **geodesic dilation** [51] is the intersection between the geodesic ball $\Omega_A(\boldsymbol{x}, \lambda)$ and a mark domain $B$:

$$\delta_A^\lambda(B) = \{\boldsymbol{x}' \in A \mid A, B \subset X^2, \boldsymbol{x} \in A, \Omega_A(\boldsymbol{x}, \lambda) \cap B \neq \emptyset\} \tag{11}$$

*Operation* 9: **reconstruction** [52] is a process of reshaping. If $f$ and $g$ are two grayscale images defined on the same domain and $f < g$, reconstruction iterates geodesic dilation $\delta_g^\lambda(f)$ until convergence:

$$\rho_g(f) = \bigvee_{\lambda > 0} \delta_g^\lambda(f) \tag{12}$$

*Operation* 10: **regional maxima** [53] extracts a domain $D_{max}(f)$ of a difference between $f$ and reconstruction $\rho_f(f)$ with power tolerance $\varepsilon$:

$$D_{\max}(f) = f - \rho_f(f - \varepsilon) \tag{13}$$

*Operation* 11: **Laplacian filter** [54] (2nd order derivative field) is the difference between the **external** ($\triangledown^+$) and **internal** ($\triangledown^-$) **gradients**:

$$\nabla^2 f = \nabla^+ f - \nabla^- f \tag{14}$$

where

$$\nabla^+ f = f \oplus g - f \tag{15}$$

$$\nabla^- f = f - f \ominus g \tag{16}$$

*Operation* 12: **Zone of Influence** (IZ) [55], [56] $Z_{x\lambda}(K_i)$ is the subset of points in $X_\lambda^2$ at a finite geodesic distance from the $i$-th intersection domain $K_i$ and closer to $K_i$ than any other $j$-th intersection $K_j$.

$$Z_{X_\lambda}(K_i) = \left\{ x \in X_\lambda^2 \,\middle|\, \begin{array}{c} d_{X_\lambda}(x, K_i) < +\infty, \\ \forall j \neq i, d_{X_\lambda}(x, K_i) < d_{X_\lambda}(x, K_j) \end{array} \right\} \tag{17}$$

where $K_i$ is the $i$-th intersection domain between $X_\lambda^2$ and $f(x)$, $K_i = f(x) \cap X_\lambda^2$.

*Operation* 13: **Skeleton by Zone of Influence** (SKIZ) [55], [56] $S(K, G)$ is the complement set of all the IZ with:

$$S(K; G) = G - \bigcup_i Z_{X_\lambda}(K_i) \tag{18}$$

### B. Watershed Segmentation

Pixels are grouped into general segments $WS$ that are projection of a 3D valleys in a field $f(\boldsymbol{x})$ onto a 2D pixel plane. Valleys can be segmented between minima and around local maxima. So, watershed segmentation is to find the local minima centers and local maxima boundaries of clusters [55], [56]. It can be linked to a problem of damming watersheds at the maxima to avoid flooding the low basin. The domains enclosed by the watersheds are the target clusters. To achieve this aim, general watershed transformation algorithm in Algorithm 1 is used [55], [56]. A 3D field $f(\boldsymbol{x})$ is cut by several angular plane $X_\lambda^2$ at level $\lambda$. The intersections between $X_\lambda^2$ and $f(\boldsymbol{x})$ are a set of domain $K$ in (17). While sweeping intensity levels from $\lambda_1$ to $\lambda_N$, the intersection domains $K$ are extended with reconstruction (12). When two $K$ contact each other, the bound is determined as SKIZ in (18), while the new intersections are IZ in (17). When reaching the maximum level $\lambda_N$, the target cluster set $W_N$ is obtained as the collection of final IZs, while the watershed set $WS$ is the complement of the clusters in the $X_\lambda^2$ at $N$. The number of level $N$ is 255 for a grayscale image. In PAS clustering, the $\lambda_N$ is the maximum of the function $f(\boldsymbol{x})$, while the choice of total level number, $N$, influences the running time of simulation.

---

**Algorithm 1** General flow of watershed segmentation

---

1: $W_1 \leftarrow \emptyset, \lambda_N \leftarrow \max f(x)$.
2: **for** $i = 1$ to $N$ **do**
3:    $m_{i+1}(f) \leftarrow$ Equation (19) at level $\lambda_i$ with (12):

$$m_{i+1}(f) = \rho_{K^{i+1}}(K^i) \tag{19}$$

   where

$$K_j = \left[X_{\lambda_i}^2 \cap f(x)\right]_j \tag{20}$$

$$K^i = \left[\bigcup_j K_j\right]_{X_{\lambda_i}^2} \tag{21}$$

4:    IZ $\leftarrow$ Equation (17).
5:    $W_i \leftarrow$ Equation (22) with

$$W_{i+1} = \left[\bigcup_j IZ_{X_{\lambda_i}}(K_j)\right] \bigcup m_{i+1} \tag{22}$$

6: **end for**
7: Watershed $WS \leftarrow$ Equation (23) as the final SKIZ with (18) at the top level $\lambda_N$

$$WS = W_N^c = X_{\lambda_N}^2 - W_N \tag{23}$$

8: Retrun $W_N$ and $WS$

---

In order to operate the watershed algorithm in Algorithm 1, power convex hill in Fig.2 needs to be transformed to valleys.

Therefore, the watershed transformation is here applied onto the Laplacian of the 2D PAS, $\triangledown^2 f$, as defined in (14). To be able to apply the water segmentation described in Algorithm 1, some pretreatment of PAS is as well as some extra steps to filter the speckles and fluctuations in Fig.2(b) to avoid over-segmentation (i.e., artificially creating a too large number of clusters), as described in Algorithm 2. The step 1 removes speckles and partly smoothen the fluctuation caused by thermal noise. In step 2, the gradient field is obtained by Laplacian operation and its contrast is enhanced in step 3. To mitigate the over-segmentation that typically occurs near the edge of clusters where the fluctuation of the Laplacian of the field is large, foreground and background markers are introduced. The foreground markers are the local maxima of the original PAS with (13), while the background markers are the curves equidistant to the clusters in the foreground with (8).

---

**Algorithm 2** Flow of watershed segmentation solving over-segmentation problem

1: **Despeckling and smoothing**: remove isolated speckles with a combination of opening (6) and closing (7); smoothen the noised PAS with reconstruction (12) and average filtering.
2: **Extract gradient field**: calculate the curvature with the Laplacian filter $\triangledown^2 f$ (14).
3: **Enhance the contrast of gradient field**: enhance contrast with closing (7); and reconstruction (12) and average filtering.
4: **Extract the marks of foreground**: get locations of $M$ regional power maxima of the foreground as centroid positions using (13) on the PAS within the interval of gradient level as tolerance $\varepsilon = \lambda_i - \lambda_{i-1}$.
5: **Extract the marks of background**: the marks are the curves equidistant to the domain with curvature in the PAS, which are the negative part of the Laplacian gradient field. The distances of every point in the background marker are calculated using (8) on the PAS directly.
6: **Group clusters**: combining the Laplacian field, marker of foreground and marker of background, operate the watershed segmentation with watershed transformation Algorithm 1.

---

Fig.3 and Fig.4 show the process of obtaining the Laplacian gradient field. The result of despeckling and noise smoothing operations (step 1 in Algorithm 2) applied on the PAS in Fig.2(b) is shown in Fig.3(a). The speckles are well removed. Furthermore the PAS mean SNR has increased to 22.05 dB with a reduced standard deviation of 2.75 dB showing that the noise has been smoothened. Then, the gradient field obtained by Laplacian filtering (step 2 in Algorithm 2)) is shown in Fig.3(b). Expect for the LOS cluster, the edges of the other clusters are fuzzy and this may jeopardize the watershed transformation. Consequently, closing and reconstruction operations (step 3 in Algorithm 2) enhance the contrast and most of the valleys in the fields exhibits clear edges in Fig.3(c).

Fig.4 shows the results of the three remaining steps in Algorithm 2. Steps 4 and 5 are specifically introduced to avoid over-segmentation caused by the gradient field fast fluctuation in the vicinity of cluster edges. They introduce a constraint on IZ operation in the general watershed transformation of Algorithm 1. The step 4 creates markers of the illuminated foreground as local field maxima as indicated in Fig.4(a). The number $M$ of local maxima is automatically found thanks to operation 13 and therefore does not need to be set a priori. The step 5 creates background markers with as shown watershed curves of the Euclidean distance field determined with (8) in Fig.4(b). The foreground and background markers are the boundary of IZ operation domain: only the elements between the foreground and background markers are effective to calculate IZ in (17). The clusters are finally obtained using watershed transformation in step 6 and are shown in Fig.4(c).

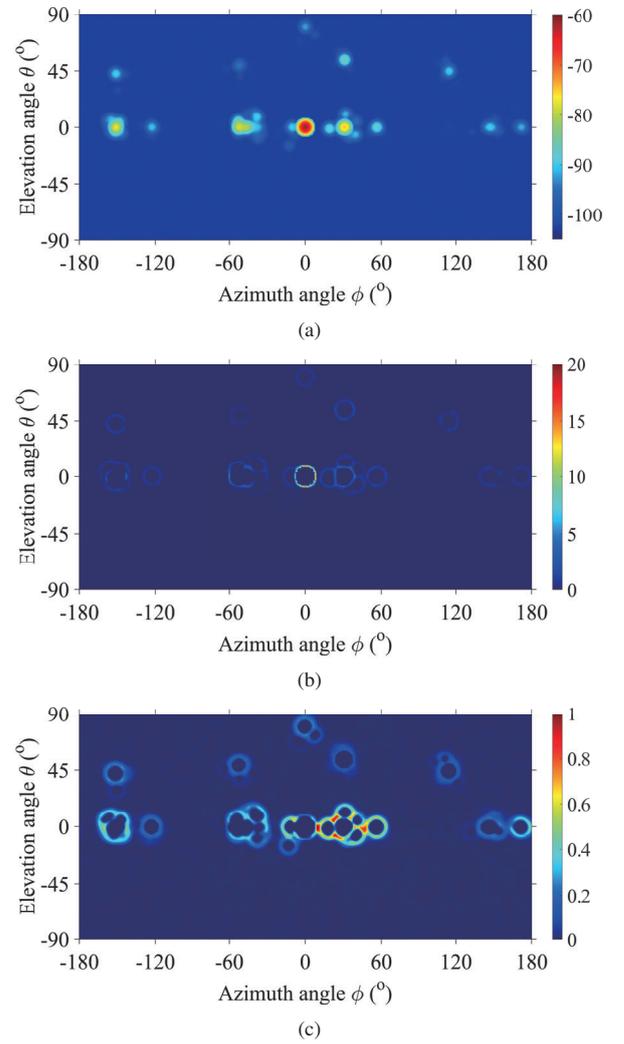

Fig. 3. Process to obtain Laplacian gradient field (using the PAS example of Fig.1): (a) PAS after despeckling and noise smoothing; (b) gradient field of PAS obtained by Laplacian operation; (c) Laplacian field enhanced contrast.

The watershed transformation uses the combination of gradient field, foreground, and background. When two intersection domains between $X_\lambda^2$ and $f(\boldsymbol{x})$, namely, $K_i$, for the valley marked with the foreground marker and $K_j$, for the valley outside the marked valley but enclosed by the background marker, contact with each other, two dual IZs are created by $K_i$ and $K_j$ with (17). All the marked valleys in the gradient field are segmented, while the valleys unmarked are neglected. Comparing the original PAS in Fig.2(a) with the clusters in Fig. 4(c), it can be qualitatively observed that Algorithm 2 meets the original expectation of the proposed clustering approach. One important parameter is the shape of $g$ in (6) and (7) that influences the denoising and smoothing operation. $g$ is a $n \times n$ matrix forming an angular filter of a given shape (e.g., disk, square diamond), depending of the values of the elements of $g$ (0 or 1). In this study, a $3 \times 3$ square matrix is empirically found to perform well. Depending on parameters such as the angular step and the beamwidth, this operator may be adjusted to obtained optimal performance.

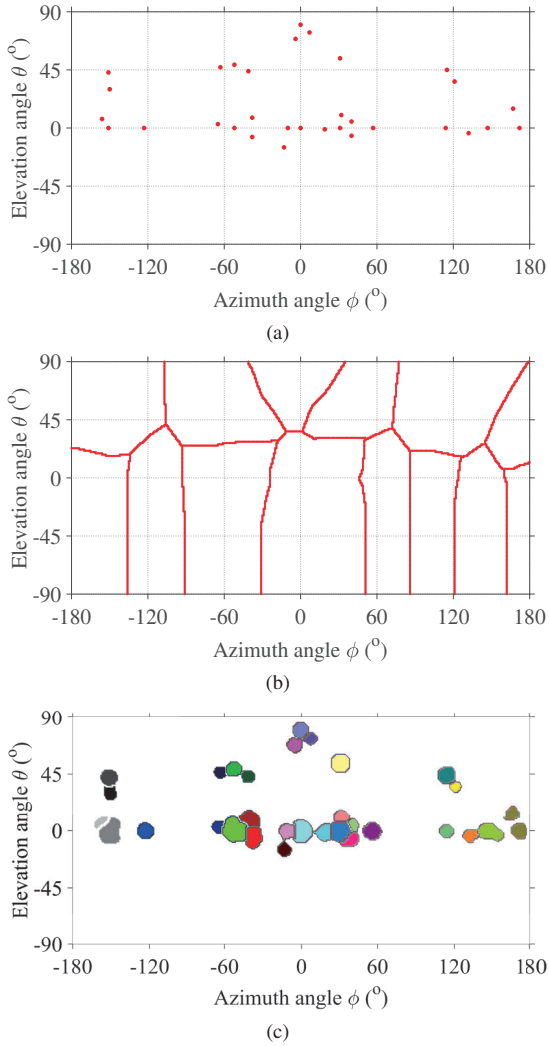

Fig. 4. Results of Algorithm 2 at different steps: (a) local maxima of original PAS as foreground markers; (b) maximum distance curves as background markers; (c) clusters marked with different colors.

### C. Clustering Comparison: modified K-Power-Means

To assess the performance of watershed, the 2D PAS are also clustered with standard iterative K-Power-Means (KPM) [31] in section IV. Furthermore, in order to investigate the influence of the pre-processing steps in the modified watershed transformation introduced in Algorithm 2, similar steps are introduced in the standard KPM as another benchmarking method, named here modified K-Power-Means (modified KPM). In particular, fixed local maxima replace the iterative searching for centroids and opening and closing operations are used to remove the speckles. Furthermore, a threshold is used to remove the background whose value is selected using Otsu's method [57]. Considering the sparsity of the millimeter-wave channel, the majority of PAS pixels represents the background rather than the clusters. Therefore, Otsu's method extracts the power value Pback of a PAS background by finding the pixel with highest probability in the intensity value histogram. The threshold $P_{thre}$ is then determined using the mean value $\mu_{SNR}$ and the standard deviation $\sigma_{SNR}$ of the SNR:

$$P_{thre} = \frac{A(B+1)}{B(A+1)} P_{back} \quad (24)$$

where

$$\begin{aligned} 10\log_{10} A &= \mu_{SNR[dB]} \\ 10\log_{10} B &= \mu_{SNR[dB]} - 3\sigma_{SNR[dB]} \end{aligned} \quad (25)$$

Here the threshold is chosen three time higher than the mean SNR by a factor equal to three times the standard deviation $\sigma_{SNR}$ in order to remove 95% of the noise fluctuation. The flowchart of the modified KPM algorithm is shown in Algorithm 3. The Multipath Component Distance (MCD) in the flow chart is the Euclidean distance used to evaluate the difference between individual multipath components. In this paper, the $i$-th parameter point is constructed with the azimuth $\phi$ and elevation $\theta$ as $(\phi_i, \theta_j)$. MCD between the $i$-th and $j$-th points is:

$$MCD_{ij} = \sqrt{(\varphi_i - \varphi_j)^2 + (\theta_i - \theta_j)^2} \quad (26)$$

**Algorithm 3** Flowchart of modified K-Power-Means algorithm
1: Remove isolated speckles with a combination of opening (6); smooth the noised PAS with reconstruction (12).
2: Extract locations local maxima power as centroid positions $c_1(0), \cdots, c_K(0)$. Remove the isolated point noise with a combination of opening (6) and closing (7), then smooth it with restructuration (12).
3: Remove the background with a threshold $P_{thre}$ with Equation (24).
4: Assign MPCs to cluster centroids and store indices $\mathcal{I}_l^{(i)}$:

$$\mathcal{I}_l^{(i)} = \arg\min_k \left\{ P_l \cdot \text{MCD}\left(x_l, c_k^{(i-1)}\right) \right\} \quad (27)$$

$$\mathcal{I}^{(i)} = \left[\mathcal{I}_1^{(i)} \ldots \mathcal{I}_L^{(i)}\right], \mathcal{C}_k^{(i)} = \underset{l}{\text{indices}}\left(\mathcal{I}_l^{(i)} = k\right) \quad (28)$$

5: Return $\mathcal{R}_k = [\mathcal{I}^{(i)}, c_k^{(i)}]$

## IV. SIMULATION VALIDATION

### A. Simulation Conditions

One thousand realizations of the IEEE 802.11ad channel model using the conference room scenario are generated to compare the performance of the watershed segmentation with KPM and Modified KPM clustering methods. Channels are obtained considering an omnidirectional Tx antenna and an Rx directional antenna beam-scanning across the 2D angular space. The maximum gain of Rx antenna is 25 dB. The scanning step of Rx antenna is $1°$ in both elevation and azimuth. The 2D PAS pixels size is thus $1° × 1°$.

### B. Qualitative Comparison between Watershed and K-Power-Means

In order to study the influence of Rx antenna radiation pattern on clustering, Fig.5 shows the result of watershed segmentation for the PAS with beamwidths of $5°$, $15°$, and $25°$. White closed curves are the labels for the identified clusters, i.e., the foreground domains. The dark blue domain is the background domain. Cluster labels clearly distinguish adjacent foreground domains. Most of the power in the foreground is gathered into clusters, and the background with weak power intensity is clearly excluded from clusters. Intuitively, the watershed segmentation achieves the two main purposes of clustering: extracting the illuminated foreground from the dark background, and distinguishing different illuminated domains. Furthermore, clustering can clearly be achieved with different beamwidths. Another important observation is that the cluster shapes are well preserved.

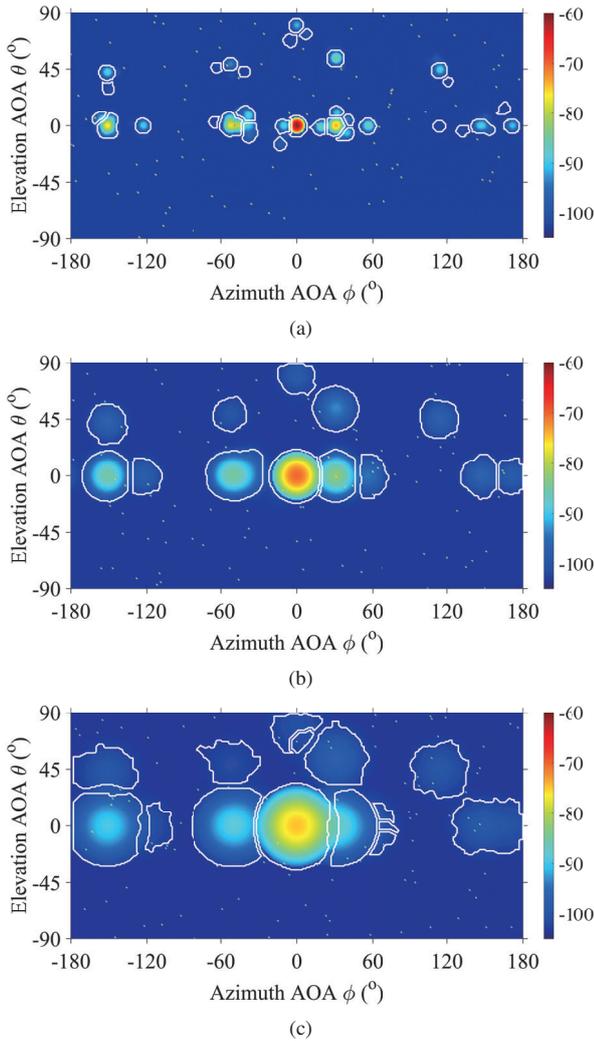

Fig. 5. Results of watershed segmentation with antenna beamwidth of (a) $5°$; (b) $15°$; (c) $25°$.

Using the standard KPM method, the entire angular space is divided into several polygons bounded by white straight lines, as shown in Fig.6. In the example in Fig.6(a), the illuminated foreground is roughly divided into large ranges, without distinguishing adjacent clusters accurately. Even worse, parts of the dark background are also enclosed into clusters. As the antenna beamwidth increases in Fig.6(b) and (c), complete high-power- intensity regions are split and arranged into different clusters. The above phenomena manifests that the standard KPM method is not sensitive to the correlation between adjacent regions. The shape of the cluster is not a polygon. So, the polygon division results in either the power leaking from a cluster into an adjacent one, or the dark background being included into a cluster.

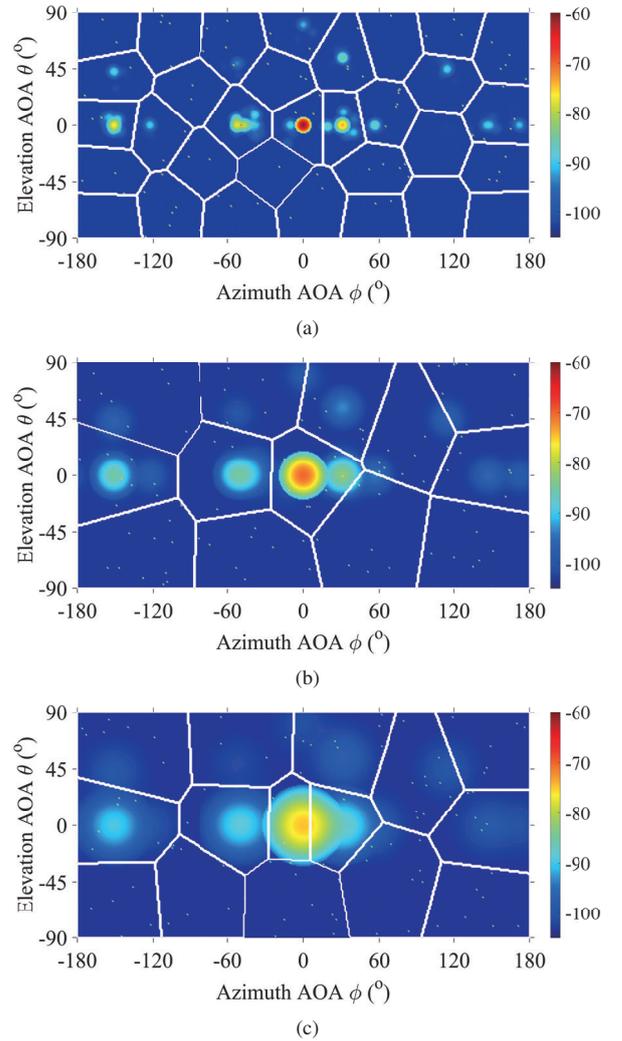

Fig. 6. Results of clustering with K-Power-Means with antenna beamwidth of (a) $5°$; (b) $15°$; (c) $25°$.

The result of modified KPM clustering is shown in Fig.7. With a narrow beamwidth of $5°$ in Fig.7(a), most clusters of the illuminated foreground can be identified, and the dark background is eliminated. However, as the beamwidth increases to $15°$, and $25°$ in Fig.7(b) and (c), respectively, the foreground markers do not improve the straight boundaries of the polygons in the standard KPM. Introduced foreground markers can find the location of certain clusters and thresholds can remove part of the background. However, part of the background is still enclosed into clusters, even in narrow beam transmission. In that case, the shapes of the clusters are only the shape of a uniform threshold instead of the individual cluster shapes. In the case of wide beam, the effect of threshold disappears: the adjacent clusters cannot be distinguished robustly.

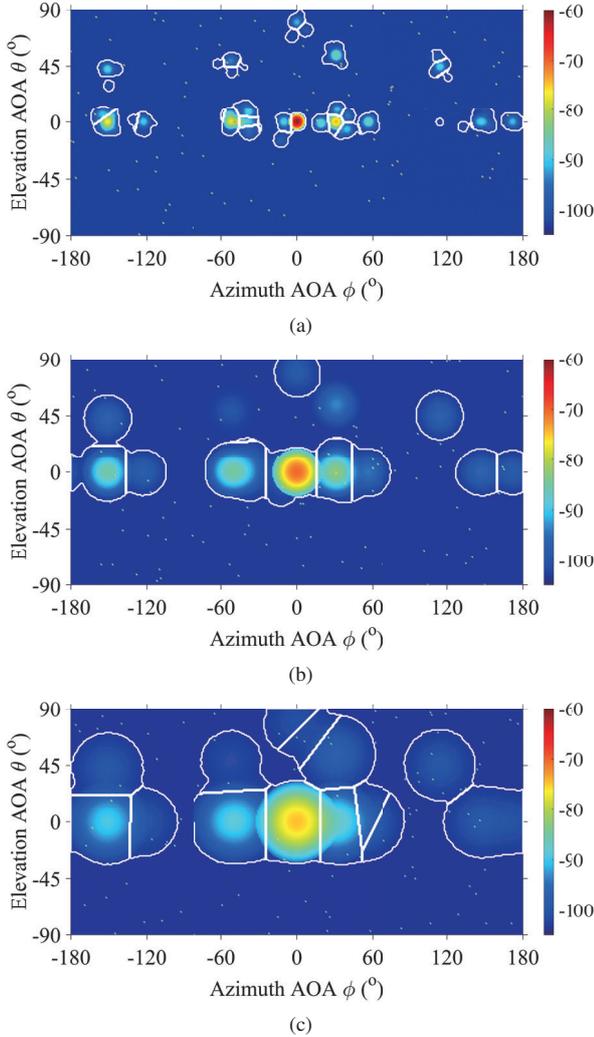

Fig. 7. Results of clustering with modified K-Power-Means with antenna beamwidth of (a) $5°$; (b) $15°$; (c) $25°$.

### C. Quantitative performance analysis of PAS Clustering

To assess the performance of the proposed clustering algorithm, several criteria are analyzed. First-of-all, the number of clusters is evaluated using the following ratio:

$$\frac{\text{Number of estimated clusters}}{\text{Number of clusters generated by channel model}} \quad (29)$$

This ratio is shown in Fig.8(a). Some discrepancies between the estimated and generated number of clusters are naturally expected (i.e., the ratio (29) is not equal to 1). Since clusters' mean angles are stochastically generated, clusters do overlap from time to time. However, it is interesting to notice that the three algorithms have the same trend. Narrow antenna beams provide higher angular resolution, so a larger number of clusters can be distinguished. As the beamwidth increases, the clusters become larger, and the corresponding number of clusters decreases. In wide beam transmission, the number of clusters provided by the watershed algorithm is closer to the number of channel clusters than the other two methods. It is also interesting to observe that all three algorithms overestimate the number of clusters when the beamwidth is narrow. Indeed, each cluster contains a few rays only, and for a given channel realization, a cluster can easily be interpreted as several clusters if the rays are angularly sparsely separated.

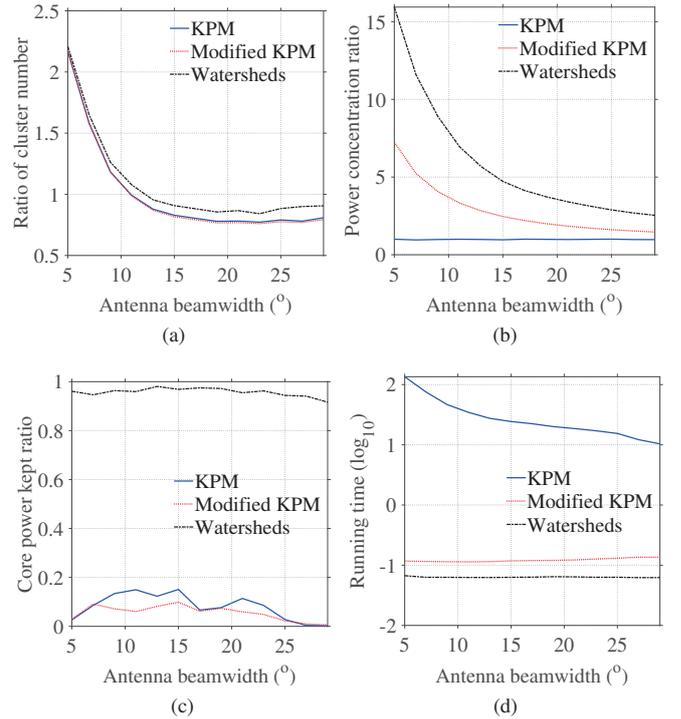

Fig. 8. Performance of clustering methods: (a) cluster number ratio, (b) power concentration ratio, (c) split cluster power ratio, and (d) running time.

The second performance criteria evaluated is the separation between the foreground and background and can be assessed by the ratio of the power density of all clusters $C$ over the power density of the whole PAS plane (after despeckling and noise removing):

$$\frac{\iint_C P(\theta,\phi)\, d\theta d\phi / \iint_C d\theta d\phi}{\iint_{PAS} P(\theta,\phi)\, d\theta d\phi / \iint_{PAS} d\theta d\phi} \quad (30)$$

The performance is presented in Fig.8(b). Since KPM cannot remove the background, the power of the entire PAS lies in the foreground, which is the sum of all clusters. Therefore, the ratio in (30) is one and is always one regardless of the beamwidth. After adding the threshold in the modified KPM method, the background is partially removed, and the ratio (30) increases. However, because background components cannot be entirely removed, some background power is also included in the cluster and the ratio is therefore not the highest. Watershed segmentation provides the most significant separation between the three algorithms. However, for wide antenna beams, the power density ratio decreases as the clustered power density is diluted into the background.

K-Power-Means and modified KPM often split clusters, which is an undesirable effect and the integrity of clusters should be therefore assessed. To assess this effect, the metric used here is the ratio of the power in the preserved illuminated clusters over the power in the damaged illuminated clusters. So, the first step is to determine a practical definition of a preserved cluster. For continuous 2D PAS map, the elements of a cluster are pixels, whose values depend on the power intensity within that cluster. A cluster is therefore a set of pixels with similar intensity compared with the neighbor domain. Because the field is continuous and derivative, the 2nd order derivative field exists. So, a cluster is enclosed by the edge of a slope. Therefore, the edge are the elements pixels at the boundary with highest 2nd order derivative. Because the intensity cluster is a continuous domain, the 2nd order derivative forms a continuous closed edge. So, the pixels inside this closed edge belong to a preserved cluster. Pixels belonging to cluster with discontinuous 2nd order derivative edge belong to damaged clusters. The following metric is subsequently defined as the ratio of the power $P(\theta, \phi)$ inside preserved clusters $C_p$ over the power inside damaged clusters $C_d$. As shown in Fig.8(c), the ratio of watershed segmentation is close to one, which means that almost all the clusters are completely preserved. In contrast, the ratios of standard KPM and modified KPM are close to zero: the two clustering algorithms split most of the clusters.

Finally, the algorithm running time is assessed. Nowadays, with high-performance ray tracing tools that exhibit reasonably realistic features, especially at millimeter waves, they can be used for channel modeling to some extent [16], [58]. This approach involves a large number of channel realizations being generated and analyzed, generating a huge volume of data [24]. Consequently, a fast clustering method is highly desirable. The simulations have been performed with a laptop (CPU 2.60 GHz, RAM 8.00 GB) and the obtained logarithmic running time is shown in Fig.8(d). When the beam is wide, the number of clusters decreases, so the required calculation time reduces. Standard KPMs takes multiple iterations to avoid local minima, so it needs a simulation time of two to three orders magnitude more than the modified KPM or watershed segmentation. While iteration is not necessary for the modified KPM, it still needs to compute the random initial centroids, which is time consuming. The watershed segmentation appears to be the fastest method among the three.

## V. MEASUREMENT VALIDATION

### A. Measurement Scenario

To verify the effectiveness of the angular clustering method, an experimental validation is conducted in a laboratory environment at Sorbonne University whose floor plan is illustrated in Fig.9. The size of the room is approximately 10.25 m × 7.52 m. The distance between the ground and the ceiling is 2.93 m. Measurements are randomly implemented in the zones which are marked as closed circles in Fig.9. Both Tx and Rx are in the same zone for a given set of experiments with distance between Tx and Rx ranging from 0.5 to 2.5 m. 100 PAS samples are measured.

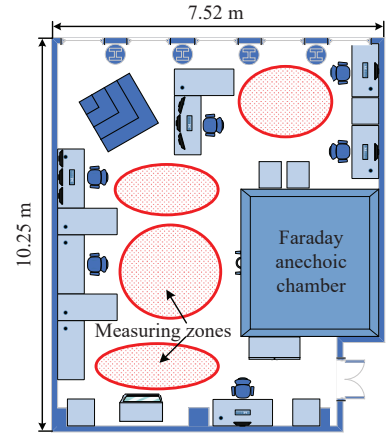

Fig. 9. Floor plan of the measuring scenario.

### B. Measurement System

The measurement set-up aims at emulating a beam training strategy as shown in Fig.10. The Tx antenna is an omni-directional dipole antenna with 2 dB gain, while the Rx antenna is a directional horn antenna with a 24 dB gain. The beam training strategy in Fig.1(a) is achieved with Rx angular scanning in vertical and horizonal directions by an azimuth motor and an elevation motor with a $5°$ angular step in both directions. The propagation channel is measured with a VNA. The set-up parameters are listed in Table I and are a tradeoff between performance and measurement time. A single PAS measurement takes approximately 3.5 hours).

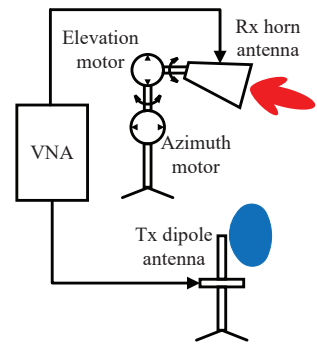

Fig. 10. Schematic of the experimental set-up.

### C. Measurement Results

An example of measured PAS is shown in Fig.11 along with clustering results of the three methods. Although not as good as in simulation, it can still be visually observed that watershed transformation grouped decently the clustered pixels in Fig.11(a). The original K-Power-Means still fail to cluster the pixels as seen in Fig.11(b). Similarly to simulations, modified KPM outperforms K-Power-Means and can cluster the pixels as shown in Fig.11(c). The overall lower performance compared with simulations is mainly due to the lower angular resolution of $5°$ step (few experimental attempts of 1°-step-size measurements have confirmed this hypothesis but the measurement duration becomes then too prohibitive: 3 days for a single PAS).

TABLE I
THE PARAMETERS OF THE PURPOSED MEASUREMENT SYSTEM

| Bandwidth | 8.64 GHz |
|---|---|
| Related time resolution | 0.12 ns |
| Frequency sample number | 752 |
| Frequency resolution | 11.5 MHz |
| Transmit power | 4 dBm |
| Noise level | -100 dBm |
| Dynamic range | 103 dB |
| Noise fluctuation of $S_{11}$ | 0.01 dB |
| Rx beam width (E/H plane) | 10.1° / 13.1° |
| Tx beam width (E/H plane) | 360° / 60° |
| Tx antenna gain | 2 dB |
| Rx antenna gain | 24 dB |
| Sampling range in azimuth | [-180°, 180°] |
| Sampling range in elevation | [-45°, 90°] |
| Angular sampling interval | 5° |

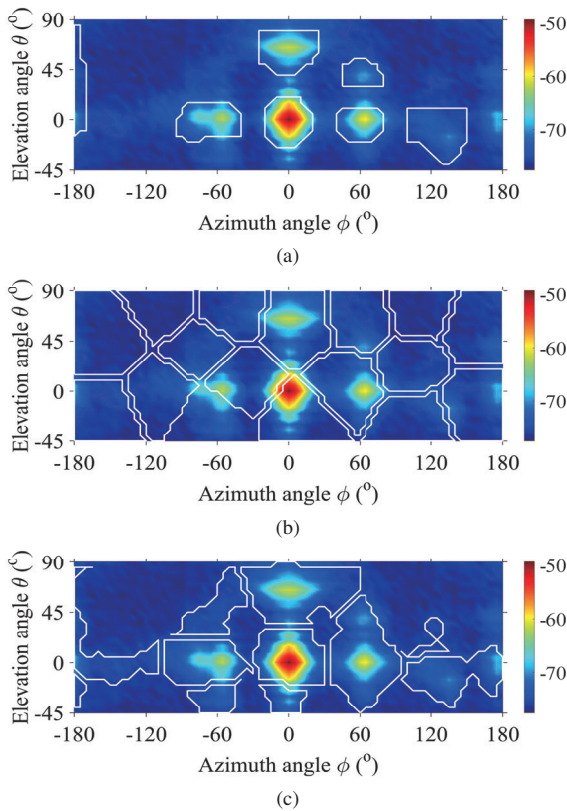

Fig. 11. Example of measured PAS in dB with: (a) watershed segmentation-based clusters, (b) KPM-based clusters and (c) modified KPM-based clusters.

The quantitative performance of clustering methods is shown in Table II. Similar to simulations, watershed segmentation still concentrates more energy as the power concentration ratio of 5.5, while the ratios of other two methods are much lower. 35.6% of the clusters are preserved with watershed transformation which is much higher than the 3.7% for modified KPM. Original K-Power-Means cannot preserve clusters at all. Watershed segmentation (0.035 s) runs little faster than modified KPM (0.067 s) and much faster than K-Power-Means (1.58 s). In summary, the result of measurement validates that watershed segmentation outperforms the other two methods.

TABLE II
PERFORMANCE OF CLUSTERING METHODS FOR MEASURED PAS

| Performance | watershed | K-Power-Means | Modified KPM |
|---|---|---|---|
| Power concentration ratio | 5.5 | 1 | 1.28 |
| Split cluster power ratio | 0.356 | 0.0000 | 0.037 |
| Running time (s) | 0.035 | 1.58 | 0.067 |

## VI. CONCLUSION

In this paper, a method based on image processing is proposed to cluster two-dimensional angular channel representations. In particular, quasi-continuous power angular spectrum maps obtained by beam-steering in azimuth and elevation are used as gray-scale images onto which clustering is performed. It is shown that watershed transformation is more suitable than classical techniques to extract illuminated clusters from the dark background and to separate adjacent clusters in these 2D maps. Furthermore, the proposed approach does preserve the shapes of clusters, which is a key criterion for performing accurate channel angular modeling. Using results obtained from 1000 realizations of the IEEE 802.11ad channel at 60 GHz, it has been shown that the proposed method significantly outperforms K-Power-Means-based algorithm in terms of identified with respect to actual number of clusters, total channel power captured within clusters with respect to background, not splitting identified clusters, and computational resources. The method has also been validated with angular (both elevation and azimuth) channel measurements conducted in an indoor scenario at 60 GHz using mechanical beam-steering.

Since the proposed approach operates on a power angular spectrum averaged over excess delay, it does not distinguish different time clusters with similar AoA. While the time dimension could be treated separately within already identified angular clusters, an interesting perspective consists in extending the proposed method to three-dimensional channel representations in order to achieve time-space clustering.

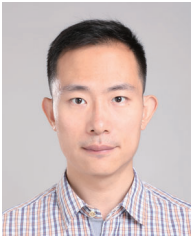

**Pengfei Lyu** received his B.S. degree from Yanshan University, China, in 2007, and two M.S. degrees from Beijing University of Technology, China, in 2010 and Institut National des Sciences Appliquées Lyon, France, in 2011, respectively. He received his Ph.D. degree from Sorbonne Université, France, in 2020. Since 2012, he has been an assistant professor with Institute of Microelectronics, Chinese Academy of Sciences, where he has been involved with design of antenna and microwave circuit, modeling of radio propagation and scattering, and numerical computation of electromagnetic wave.

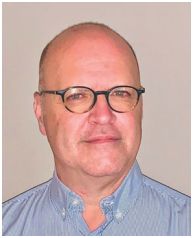

**Aziz Benlarbi-Delaï** received the PhD in Electrical Engineering and The Habilitation à Diriger des Recherches (HDR) ès Sciences Physiques from the University of Lille I in 1992 and 2002. From 1992 to 2006 he was Assistant Professor at this University and was mainly involved, as researcher at the Institut of Electronic Microelectronic and Nanotechnology (IEMN), in the field of microwave and millimetre wave devices and systems for communication and localisation. He also investigated others fields of research dealing with ultra-fast sampling devices, plasmonic structures and micro and nano technology.

He is currently a full time Professor in Electrical Engineering at Sorbonne University and head of the Electronics department of GeePs Laboratory. Former, he was director of the laboratory of Electronics and Electromagnetism (L2E) and a member of the Strategic Committee of the Labex SMART. His main research activities deal with channel modelling for wireless systems and is now involved in neuromorphic design for ultra-low power systems.

He is the author of 130 publications and communications and the holder of two patents. He participates to several Technical Program Committee of international conferences.

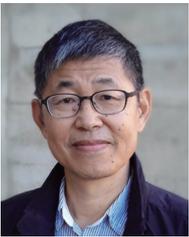

**Zhuoxiang Ren** is a full professor at the Sorbonne University in France and conducts his research at Group of Electrical Engineering Paris (GeePs). He graduated from Huazhong University of Science and Technology in China in 1982 and received his PhD from the Insititut Polytechnique de Toulouse in France in 1985. His working experience includes both academic and industrial research and development in France and in USA.

Prof. Ren has published over 200 refereed international journal and conference papers and is a co-author of two book chapters. He has been awarded a bronze medal of CNRS in France in 1996. His research interests include numerical methods for computational electromagnetics and multiphysics, modeling and simulation of electrical and electronic devices and systems.

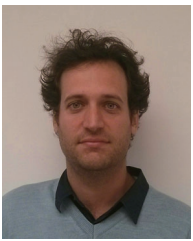

**Julien Sarrazin** received his Master and PhD degrees from the University of Nantes in France, in 2005 and 2008 respectively. In 2009 and 2010, he worked with the BK Birla Institute of Technology of Pilani, India. In 2011 and 2012, he was a research engineer at Telecom ParisTech in Paris. Since September 2012, he is an Associate Professor at Sorbonne Université in Paris, where he is currently working in the GeePs research institute (Group of Electrical Engineering of Paris) in the field of spatial data focusing, antenna design, channel modeling, and physical layer security.